\documentstyle[twocolumn,pra,aps]{revtex}

\newcommand{\be}{\begin{equation}}
\newcommand{\ee}{\end{equation}}
\newcommand{\ba}{\begin{eqnarray}}
\newcommand{\ea}{\end{eqnarray}}

\draft

\begin{document}
\draft
\title{Ideal gas in nonextensive optimal Lagrange multipliers formalism}
\author{Sumiyoshi Abe$^{1}$, S. Mart\'{\i }nez$^{2,\,3}$, F. Pennini$^{2}$,
and A.
Plastino${^{2,\,3}}$}
\address{$^1$ College of Science and Tehcnology, Nihon University, Funabashi,\\
Chiba 274-8501, Japan}
\address{$^2$ Physics Department, National University La Plata, C.C. 727,\\
1900 La Plata, Argentina}
\address{$^3$ Argentina National Research Council (CONICET)}
\maketitle

\begin{abstract}
Based on the prescription termed {\it the optimal
Lagrange multipliers} formalism for extremizing the Tsallis
entropy indexed by $q$, it is shown that key aspects of the treatment
of the ideal gas problem are identical in both the
nonextensive $q \ne 1$ and extensive $q=1$ cases.
 \vskip 3mm

\vspace{1 cm}
\end{abstract}

\section{Introduction}

The purpose of this work is to discuss the classical ideal gas
problem in Tsallis thermostatistics within the framework
of the method of optimal Lagrange multipliers (OLM) recently
proposed in Ref. \cite{olm}.

The ideal gas problem in the normalized Tsallis thermostatistics
has recently been exhaustively discussed in Ref. \cite{abe1}.
Correlations induced by Tsallis' nonextensivity were analyzed in
Ref. \cite{abe2}. An interesting relation, analogous to that
yielding the mean energy of the (ordinary) ideal gas was thereby
found. However, the ensuing results still depend upon Tsallis'
nonextensivity index $q$, even if one deals with a purely
classical system. Here, we show that more insights can be gained
into the problem if i) one takes advantage of a degree of freedom
implicit in Tsallis' formalism, i.e., choice of the entropy
constant $k_T$, playing the role of Boltzmann's constant $k_B$,
  and ii) uses the OLM technique in Ref. \cite{olm}.

Tsallis' thermostatistics [4-18] is by now known to offer a
nonextensive generalization of traditional Boltzmann-Gibbs
statistical mechanics. A key ingredient in this formalism is the
introduction of a particular definition of expectation value
termed the normalized $q$-expectation value \cite{mendes}.
Actually, during the last ten years before the work  in Ref.
\cite{mendes}, several proposals have been made regarding
definition of expectation value \cite{review}. No matter what
definition one chooses, ordinary Boltzmann-Gibbs results are
always reproduced in the extensive limit $q\rightarrow 1$.

Tsallis thermostatistics involves extremization of Tsallis' entropy

\begin{equation}
\frac{S_{q}}{k_T}=\frac{1-\int d{\bf x\ }f^{q}({\bf x})}{q-1},
\label{entropia}
\end{equation}
by recourse to Lagrange's constrained variational technique. Here,
${\bf x}$ is a phase space element
  ($N$ particles in a $D$-dimensional space),
  $q \in \Re$ is Tsallis' nonextensivity index, $f$
stands for any normalized probability density, and
$k_T$ is the entropy constant that is akin to the
celebrated Boltzmann constant appearing in traditional
statistical mechanics.

The formalism of Ref. \cite{mendes} gives rise to a
non-diagonal form of the Hessian associated with
examining the extremum structure of the entropy and
free energy through the Legendre transform procedure.
In Ref. \cite{olm}, a new approach has been advanced,
in which the Hessian is diagonal. This approach was
shown to enormously facilitate ascertaining what kind
of extrema the Lagrange technique yields.

A quite interesting fact is to be emphasized concerning
the entropy constant $k_T$, which is usually identified
simply with Boltzmann's $k_{B}$ in the literature: the
only certified fact one can be sure of is ``$k_T
\rightarrow k_{B}$ for $q\rightarrow 1$''
\cite{review}, which entails that {\it there is room to
choose $k_T=k(q)$ in a suitable way}. It is seen
\cite{olm} that if one chooses (with $\bar{Z}_{q}$ the
generalized partition function)

\begin{equation}
k_T=k_{B}\bar{Z}_{q}^{q-1}, \label{k}
\end{equation}
in conjunction with the OLM formalism \cite{olm}, the
classical harmonic oscillator determines a specific
heat $C_{q}=k_{B}$, so that {\it the ordinary
Boltzmann-Gibbs result arises without invoking the
limit} $q\rightarrow 1$. In addition, the OLM treatment
is able to reproduce the zero-th law of thermodynamics,
a goal that had previously eluded
Tsallis-thermostatistics practitioners \cite{ley0}. One
is then tempted to conjecture that many other ordinary
results can be reproduced by Tsallis
thermostatistics without invoking the limit
$q\rightarrow 1$. In this paper, we revisit the ideal
gas problem in Tsallis thermostatistics with such a
goal in mind. More precisely, we perform an OLM
treatment of the problem following the canonical
ensemble structure.
    We will show that the ordinary expression for the
    mean value of the energy is reproduced by Tsallis thermostatistics
    without the need of going to the limit
    $q\rightarrow 1$, in contrast to the previous treatment of this problem.

    It is worth noting here that, as shown in Ref.
    \cite{plagibbs}, the usual thermodynamic limit $N \rightarrow \infty$
     implies the limit
$q\rightarrow 1$. There, the Tsallis probability distribution is
deduced in a way that mimics Gibbs' celebrated derivation of the
canonical distribution for a system in contact with a heat bath
\cite{plagibbs}. It was shown that Tsallis' distribution naturally
arises for {\it finite heat baths}, the nonextensivity index $q$
being related to the particle number $N$ that characterizes the
bath. Gibbs' canonical distribution, instead, results for infinite
heat baths \cite{plagibbs}. This leads one to conclude that $q$
goes over to unity in the limit $N \rightarrow \infty$. This
result was found employing the unnormalized expectation values of
Curado and Tsallis (see Ref. \cite{review}). Analogously, if one uses
the normalized expectation values of Ref. \cite{mendes}, a similar
argument leads to
\be
q=\frac{D N-4}{D N-2} \ee for the ideal gas in $D$ dimensions. Since
the limit $N \rightarrow \infty$ corresponds to $q \rightarrow 1$,
the same ordinary result is obtained from the nonextensive formalisms with
various definitions of generalized expectation value (both normalized and
unnormalized) in such a limit.

\section{Brief review OLM formalism}

A general classical treatment requires consideration of the
probability density $f({\bf x})$ that maximizes Tsallis' entropy,
subject to the foreknowledge of the generalized expectation values
of certain physical quantities.

Tsallis' probability distribution \cite{mendes} is
obtained by following the well known MaxEnt principle
\cite{katz}. The Tsallis-Mendes-Plastino variational
treatment \cite{mendes}
  involves a set of Lagrange multipliers $%
\lambda_j$.
   The OLM technique developed in Ref. \cite{olm} pursues
an alternative path that involves a different set of
Lagrange multipliers, say, $\lambda^{\prime}_j$: one
maximizes Tsallis entropy $S_q$ in Eq. (\ref{entropia})
\cite{t01,t1,t2}, subject to the {\it modified}
constraints (``centered" generalized expectation
values) \cite{olm,t01}:

\begin{eqnarray}
\int d{\bf x\ }f({\bf x})-1 &=&0, \\
\int d{\bf x}\ f({\bf x})^{q}\left( O_{j}({\bf x})-\left\langle \left\langle
O_{j}\right\rangle \right\rangle _{q}\right) &=&0,  \label{vinculos}
\end{eqnarray}
where $O_{j}({\bf x})$ ($j=1, 2, \ldots , M$) denote the $M$ relevant dynamical
quantity (the observation level \cite{aleman}). In the above,
$\left\langle \left\langle O_{j}\right\rangle \right\rangle _{q}$
are defined by \cite{mendes}

\begin{equation}
\left\langle \left\langle O_{j}\right\rangle \right\rangle _{q}=\frac{\int d%
{\bf x\ }f^{q}({\bf x})O_{j}({\bf x})}{\int d{\bf x\ }f^{q}({\bf x})},
\label{gener}
\end{equation}
which are assumed to be {\it a priori} known. (The procedure
given in Ref. \cite{mendes}  employs non-centered expectation
values.) The resulting probability distribution reads \cite{olm}

\begin{equation}
f({\bf x})=\bar{Z}_{q}^{-1}\left[ 1-(1-q)\sum_{j}^{M}\,\lambda _{j}^{\prime
}\left( O_{j}({\bf x})-\left\langle \left\langle O_{j}\right\rangle
\right\rangle _{q}\right) \right] ^{\frac{1}{1-q}},  \label{rho}
\end{equation}
where $\bar{Z}_{q}$ stands for the generalized partition function

\begin{equation}
\bar{Z}_{q}=\int d{\bf x}\left[ 1-(1-q)\sum_{j}^{M}\lambda _{j}^{\prime
}\left( O_{j}({\bf x})-\left\langle \left\langle O_{j}\right\rangle
\right\rangle _{q}\right) \right] ^{\frac{1}{1-q}}.  \label{Zqp}
\end{equation}

Although the procedure originally devised in Ref.
\cite{mendes} overcomes some problems posed by the old,
unnormalized way of evaluating Tsallis' generalized
expectation values \cite{mendes,pennini}, it yields
probability distributions that are self-referential. The resulting
distribution includes the integral of the $q$th power of the
distribution itself. This fact entails difficulties in numerical model
calculations, for example. The complementary OLM treatment of
Ref. \cite{olm} surmounts such difficulties. The above-mentioned
self-reference problem does not arise in Eq. (\ref{rho}).

It is shown in Ref. \cite{olm} that the Lagrange multipliers
$\lambda _j$ of the Tsallis-Mendes-Plastino
procedure \cite{mendes} and the corresponding
$\lambda _j^{\prime }$ in OLM are connected to each other as follows:

\begin{equation}
\lambda _{j}^{\prime }=\frac{\lambda _{j}}{\int d{\bf x\ }f^{q}({\bf x})}.
\label{conectl}
\end{equation}
However, the genuine Lagrange multipliers are $\lambda _{j}^{\prime }$ in OLM.

The probability density appearing in Eq. (\ref{conectl}) is the
one that maximizes the entropy $S_q$ which can be expressed in the
alternative form \cite{olm}

\begin{equation}
S_{q}=k_T \frac{\bar{Z}_{q}^{q-1}-1}{q-1}\int d{\bf x\
}f^{q}({\bf x}).
\label{Sq}
\end{equation}

Also, the identical relation

\begin{equation}
\int d{\bf x\ }f^{q}({\bf x})=\bar{Z}_{q}^{1-q}  \label{relac1}
\end{equation}
holds, from which it follows \cite{olm} that

\begin{equation}
S_{q}=k_T\;{\rm \ln }_{q}\bar{Z}_{q}, \label{S2}
\end{equation}
where ${\rm \ln }_{q}x=(1-x^{1-q})/(q-1).$

Eq. (\ref{relac1}) allows us to rewrite Eq. (\ref{conectl}) as

\begin{equation}  \label{lambda'}
\lambda_j^{\prime }= \frac{\lambda_j}{\bar{Z}_{q}^{1-q}}.
\end{equation}
However, we again stress that the basic variables in
OLM are $\lambda_j^{\prime }$.

Now, following Ref. \cite{mendes}, we define

\begin{equation}
\ln _{q}Z_{q}^{\prime }={\rm \ln }_{q}\bar{Z}_{q}-\sum_{j}\lambda
_{j}^{\prime }\ \left\langle \left\langle O_{j}\right\rangle \right\rangle
_{q}.  \label{lnqz'}
\end{equation}
Introducing

\begin{equation}
k^{\prime }=k_T\;\bar{Z}_{q}^{1-q} \label{k'}
\end{equation}
as in Ref. \cite{olm}, we straightforwardly obtain

\begin{eqnarray}
\frac{\partial S_{q}}{\partial \left\langle \left\langle O_{j}\right\rangle
\right\rangle _{q}} &=&k^{\prime }\lambda _{j}^{\prime },  \label{termo1} \\
\frac{\partial }{\partial \lambda _{j}^{\prime }}\left( \ln
_{q}Z_{q}^{\prime }\right) &=&-\left\langle \left\langle O_{j}\right\rangle
\right\rangle _{q}.  \label{termo2}
\end{eqnarray}

Eqs. (\ref{termo1}) and (\ref{termo2}) constitute
the basic information-theoretic relations, on which statistical
mechanics can be built {\em \`a la} Jaynes \cite{katz}. Here, one should
remind the well known fact that in reconstructing statistical
mechanics based on information theory, Jaynes could remove the concept
of ensemble \cite{katz}. In this
sense, it is appropriate to emphasize that both the
OLM \cite{olm} and Tsallis-Mendes-Plastino
\cite{mendes} formalisms employ identical {\it a priori}
information, so that they are physically equivalent.

Notice that $k^{\prime}$, as defined by Eq. (\ref{k'}), obeys the condition $%
k^{\prime} \rightarrow k_B$ as $q \rightarrow 1$. It is
a condition that this constant must necessarily
fulfill. (See Ref. \cite{t01}). Notice that if one
makes use of the possibility of choosing $k_T$ as in
(\ref{k}), one obtains, on account of (\ref{k'}),
\be
k^{\prime} = k_B, \ee i.e., the classical results are
obtained without going to the limit $q \rightarrow 1$,
as stated in the introduction.

  Moreover, another interesting result obtained from OLM
is \cite{olm}

\begin{equation}
k^{\prime }\lambda _j^{\prime }=k_T \lambda _j,
\label{rel}
\end{equation}
which entails that the intensive variables are the same in both the OLM
   and Tsallis-Mendes-Plastino  formalisms.

As a special instance of Eqs. (\ref{termo1}), (\ref{termo2}) and
(\ref{rel}), let us discuss the canonical ensemble. In this case,
only a single constraint regarding the system Hamiltonian is
considered. Writing the associated Lagrange multiplier as $\beta
^{\prime }$ and the generalized internal energy as $U_q$,

\begin{eqnarray}
\frac{\partial S}{\partial U_q}=k^{\prime }\beta ^{\prime
}&=&k_T
\beta =\frac {1}{T},  \label{C1} \\ \frac \partial {\partial \beta
^{\prime }}\left( \ln _qZ_q^{\prime }\right) &=&-U_q, \label{C2}
\end{eqnarray}
where
\begin{equation}
\ln _qZ_q^{\prime }=\ln _q\bar{Z}_q-\beta ^{\prime }U_q.  \label{lnqz'can}
\end{equation}

    The temperature $T$ in Eq. (\ref{C1}) is the same as that in the
Tsallis-Mendes-Plastino  formalism.

\label{NTT} \section{Ideal gas in Tsallis-Mendes-Plastino formalism}

The classical ideal gas in $D$-dimensional space in the
Tsallis-Mendes-Plastino formalism has
been considered in Refs. \cite{abe1,abe2}. For comparison, we recapitulate
here some of
its main results in the case $0<q<1$. The Hamiltonian reads

\begin{equation}
H({\bf P})=\sum_{i=1}^{N}\frac{{\bf p}_{i}^{2}}{2m},  \label{H}
\end{equation}
where $m$ is the particle mass, $N$ the particle number
and ${\bf p}_{i}$ the momentum of the $i$th particle. We are
writing the $N$-particle momenta collectively as ${\bf P}=
({\bf p}_{1}, {\bf p}_{2}, \ldots ,{\bf p}_{N})$.
One extremizes the entropy in Eq.
(\ref{entropia}), subject to the constraints \cite{mendes}

\begin{eqnarray}
\int d\Omega f({\bf P}) &=&1, \\
\frac{1}{c}\int d\Omega f({\bf P})^{q}H({\bf P}) &=&U_{q},  \label{vinc}
\end{eqnarray}
where $d\Omega
=(1/(N!h^{DN}))\prod_{i=1}^{N}d{\bf q}_{i}d{\bf p}_{i}$, with
$h$ the
linear dimension (i.e. the size) of the elementary cell
in phase space, and $\int
\prod_{i=1}^{N}d{\bf q}_{i}=V^{N}$ with the spatial volume $V$. We have
also introduced the quantity $c$

\begin{equation}
c= \int d\Omega f({\bf P})^q.
\end{equation}

The equilibrium probability density is found to be
\begin{equation} \label{equi}
f({\bf P})=\frac{1}{\tilde{Z}_{q}}\left[ 1-(1-q)(\beta
/c)(H({\bf P})-U_{q})\right] ^{\frac{1}{1-q}},
\end{equation}
where the generalized partition function is
\begin{equation}
\tilde{Z}_{q}=\int d\Omega \left[ 1-(1-q)(\beta /c)(H({\bf
P})-U_{q})\right] ^{%
\frac{1}{1-q}},
\end{equation}
and we have

\begin{equation}
c=[\tilde{Z}_q]^{1-q}.
\end{equation}

In the above, $\beta$ is the Lagrange multiplier associated with
the constraint in Eq. (\ref{vinc}). Notice that Eq. (\ref{equi}) is
indeed a self-referential
expression through $c$. In the present circumstances,
however, this problem can be overcome, since a
straightforward mathematical manipulation yields

\begin{equation}
\frac{\beta U_q}{c}=\frac{DN}{2},
\end{equation}
with $c$ given by

\begin{eqnarray}
c &=&\left\{ \frac{\Gamma \left( \frac{1}{1-q}\right) }{\Gamma \left( \frac{1%
}{1-q}+\frac{DN}{2}\right) }\frac{V^{N}}{N!h^{DN}}\left[ \frac{2\pi m}{%
(1-q)\beta }\right] ^{\frac{DN}{2}}\right.  \nonumber \\
&&\left. \left[ 1+(1-q)\frac{DN}{2}\right] ^{q(1-q)+DN/2}\right\}
^{\frac{2(1-q)}{2-(1-q)DN}}.
\end{eqnarray}

To discuss the statistical properties of the particle
energies, the $i$th and $j$th single-particle
Hamiltonians may be considered, namely,
$H_{i}={\bf p}_{i}^{2}/2m$ and $H_{j}={\bf p}_{j}^{2}/2m$,
respectively. Their generalized variance, covariance
and correlation coefficient are defined by \cite{abe2}

\begin{eqnarray}
(\Delta _qH_i)^2&=&\langle \langle H_i{}^2 \rangle \rangle _q - \langle
\langle H_i{}\rangle \rangle_q^2,
\nonumber \\
C_q(H_i,H_j)&=&\langle \langle H_iH_j\rangle \rangle_q - \langle \langle
H_i\rangle\rangle _q\langle \langle
H_j\rangle \rangle_q,  \label{statistical} \\
\rho(H_i,H_j)&=&\frac{C_q(H_i,H_j)}{\sqrt{(\Delta _qH_i)^2(\Delta _qH_j)^2}},
\nonumber
\end{eqnarray}
respectively. A straightforward calculation shows that these quantities are
given by

\begin{eqnarray}
(\Delta _qH_i)^2&=& \frac{c^2}{2 \beta^2}\, \frac{2D+(1-q)D^2 (N-1)}{%
4-2q-(1-q)DN},  \nonumber \\
C_q(H_i,H_j)&=&- \frac{c^2}{2 \beta^2}\, \frac{(1-q)D^2}{4-2q+(1-q)DN}
\label{statval}, \\
\rho(H_i,H_j)&=&-\frac{(1-q)D}{2+(1-q)D(N-1)}.  \nonumber
\end{eqnarray}

In the limit $q \rightarrow 1$, one gets

\begin{eqnarray}
(\Delta _qH_i)^2&\rightarrow& \frac{D}{2} \left( \frac{h^2}{2 \pi m}%
\right)^2 e^{-(2+4/D)}\left( \frac{N}{V}\right)^{4/D},  \nonumber \\
C_q(H_i,H_j)&\rightarrow& 0,  \label{statlim} \\
\rho(H_i,H_j)&\rightarrow& 0.  \nonumber
\end{eqnarray}

\section{Ideal gas in OLM formalism}

We revisit here the classical ideal gas problem considered in the
previous section within the OLM framework. We extremize Eq.
(\ref {entropia}), subject now to the constraints

\begin{eqnarray}
\int d\Omega f({\bf P})-1 &=&0 \\
\int d\Omega f({\bf P})^{q}(H({\bf P})-U_{q}) &=&0,  \label{vinco}
\end{eqnarray}
with $H$ given in Eq. (\ref{H}).

The probability distribution $f({\bf P})$ reads

\begin{equation}
f({\bf P})=\frac 1{\bar{Z}_q}[1-(1-q)\beta ^{\prime }(H({\bf
P})-U_q)]^{\frac 1{1-q}%
},
\end{equation}
where
\begin{equation}
\bar{Z}_q=\int d\Omega [1-(1-q)\beta ^{\prime}(H({\bf P})-U_q)]^{\frac 1{1-q}}.
\end{equation}

In these equations, $\beta ^{\prime }$ is the Lagrange multiplier
associated with the constraint in Eq. (\ref{vinco}).

We now follow the steps indicated in Ref. \cite{abe2} and
summarized in Section \ref {NTT}. Our interest lies in
the interval $0<q<1$ again. First, define
\begin{equation}
R_{1}=\frac{\Gamma \left( \frac{2-q}{1-q}\right) }{\Gamma \left( \frac{2-q}{%
1-q}+\frac{DN}{2}\right) },
\end{equation}
and
\begin{equation}
R_{2}=\left[ \frac{2\pi mc}{(1-q)\beta ^{\prime }}\right] ^{\frac{DN}{2}%
}\left[ 1+(1-q)\frac{\beta ^{\prime }U_{q}}{c}\right] ^{\frac{1}{1-q}+\frac{%
DN}{2}}.
\end{equation}

The associated generalized partition function is given by

\begin{equation}
\bar{Z}_{q}(\beta ^{\prime })=R_{1}\,\,\frac{V^{N}}{N!\,\,h^{DN}}\,\,R_{2},
\label{zqpg}
\end{equation}
while, introducing
\begin{equation}
G_{1}=\frac{DN}{2\beta ^{\prime }\bar{Z}_{q}(\beta ^{\prime })},
\end{equation}
the generalized internal energy turns out to be
\begin{equation}
U_{q}=G_{1}\,\,R_{1}\,\,\frac{V^{N}}{N!\,h^{DN}}\,\,R_{2}.  \label{uqg}
\end{equation}

After replacing $\bar{Z}_q(\beta ^{\prime })$ by Eq. (\ref{zqpg}), one finds

\begin{equation}
U_q=\frac{DN}{2\beta ^{\prime }}.
\end{equation}

Now, taking advantage of the fact that (Cf. Eq.
(\ref{C1}))
\begin{equation}
\beta ^{\prime }=\frac 1{k_BT},
\end{equation}
one is immediately led to
\begin{equation}
U_q=\frac D2Nk_BT,
\end{equation}
i.e., to the ordinary result. Note that this result is obtained here {\it
without going to the limit} $%
q\rightarrow 1$. By following the methodology advanced
in Ref. \cite{olm}, we see that, with the
identification of Eq. (\ref{k}), Tsallis
thermostatistics is able to reproduce ordinary results
independently of the specific choice of the value of
$q$.

Now, let us discuss statistical properties of the single-particle energies.
Following again Section \ref{NTT}, we focus our attention on the single
particle Hamiltonians pertaining to the $i$th and the $j$th particles, $%
H_{i}={\bf p}_{i}^{2}/2m$ and $H_{j}={\bf p}_{j}^{2}/2m$, respectively. Their
generalized variance, covariance and correlation coefficient are defined
collectively in Eq. (
yields
\begin{eqnarray}
&(\Delta _qH_i)^2 &= \frac 1{2(\beta ^{\prime })^2}\frac{2D+(1-q)D^2(N-1)}{%
4-2q-(1-q)DN} ,  \nonumber \\
&C_q(H_i,H_j) & = -\frac 1{2(\beta ^{\prime })^2}\frac{(1-q)D^2}{4-2q-(1-q)DN%
}, \\
&\rho(H_i,H_j)&=-\frac{(1-q)D}{2+(1-q)D(N-1)},  \nonumber
\end{eqnarray}
respectively. In the limit $q\rightarrow 1$, one has

\begin{eqnarray}
&(\Delta _qH_i)^2&\rightarrow \frac D{2(\beta ^{\prime })^2},  \nonumber \\
&C_q(H_i,H_j)&\rightarrow 0, \\
&\rho(H_i,H_j)&\rightarrow 0.  \nonumber
\end{eqnarray}

We see that the limit $q\rightarrow 1,$ $C_{q}$ and $\rho $ behave as in
the Tsallis-Mendes-Plastino formalism in Ref. \cite{abe2}.
However, for $(\Delta _qH_i)^2$, we find something new:
{\it the result becomes independent of the density}. It
depends only on the temperature, as it does in the ordinary
Boltzmann-Gibbs treatment.

\section{Conclusions}

In this work, we have addressed the ideal gas problem within the framework of
Tsallis thermostatistics reformulated by the optimal Lagrange multipliers
(OLM) method advanced in Ref. \cite
{olm}. Three interesting observations have been presented:

\begin{itemize}
\item   The internal energy was found to be given by
\begin{equation}
U_{q}=\frac{D}{2}Nk_{B}T,
\end{equation}
which is the same as the ordinary result obtained by Boltzmann-Gibbs
theory. We derived this result {\it without going to the limit}
$q\rightarrow 1$. That is, it is valid for all $q$.

\item   The correlation coefficient for the particle energies become
density-independent in OLM, in contrast to that in the
Tsallis-Mendes-Plastino formalism in Ref. \cite{mendes}.

\item   The constant $k_T$ in the definition of the Tsallis entropy may
depend on $q$ and, therefore, on the number of
particles. In the Boltzmann-Gibbs canonical ensemble
theory, the limit {\it particle number $N$ going to
infinity} is always in mind, both explicitly and
implicitly. This indicates that $k_T\rightarrow
k_{B}$ and $q\rightarrow 1$ may correspond to $%
N\rightarrow \infty \,\,\approx \,\,Avogadro^{\prime }s\,\,\,number$.
\end{itemize}

\acknowledgements The financial support of the National Research Council
(CONICET) of Argentina is gratefully acknowledged. F. Pennini acknowledges
financial support from UNLP, Argentina.

\end{document}